\documentclass[twocolumn]{aastex6}

\usepackage{natbib}
\usepackage{rotating}
\usepackage{amsmath}
\usepackage{graphicx}
\usepackage{textcomp}
\usepackage{color}
\usepackage{tabularx}

\begin{document}

\title{The complex morphology of the young disk MWC 758: SPIRALS AND DUST CLUMPS AROUND A LARGE CAVITY}
\author {Y. Boehler$^1$, L. Ricci$^{1}$, E. Weaver$^1$, A. Isella$^1$, M. Benisty$^{2,3}$, J. Carpenter$^4$, C. Grady$^5$, 
Bo-Ting Shen$^6$, Ya-Wen Tang$^6$, and L. Perez$^{7,8}$}
\affil{$^1$Rice University, Department of Physics and Astronomy, Main Street, 77005 Houston, USA \\
       $^2$Unidad Mixta Internacional Franco-Chilena de Astronom\'{i}a, CNRS/INSU UMI 3386 and Departamento de 
       Astronom\'{i}a, Universidad de Chile, Casilla 36-D, Santiago, Chile \\
       $^3$Univ. Grenoble Alpes, CNRS, IPAG, 38000, Grenoble, France \\
       $^4$Joint ALMA Observatory (JAO), Alonso de Cordova 3107 Vitacura - Santiago de Chile, Chile \\
       $^5$Exoplanets and Stellar Astrophysics Laboratory, NASA Goddard Space Flight Center, Greenbelt, MD, USA \\
       $^6$Academia Sinica, Institute of Astronomy and Astrophysics, Taipei, Taiwan \\
       $^7$Max-Planck-Institute for Astronomy, Bonn, Germany \\
       $^8$Universidad de Chile, Departamento de Astronomía, Camino El Observatorio 1515, Las Condes, Santiago, Chile }
              
\begin{abstract}




We present Atacama Large Millimeter Array (ALMA) observations at an angular resolution of 0.1\arcsec-0.2\arcsec\ of the disk 
surrounding the young Herbig Ae star MWC~758. The data consist of images of the dust continuum emission recorded at 0.88 
millimeter, as well as images of the $^{13}$CO and C$^{18}$O J = 3-2 emission lines. The dust continuum emission is characterized 
by a large cavity of roughly 40 au in radius which might contain a mildly inner warped disk. The outer disk features two bright 
emission clumps at radii of $\sim$ 47 and 82 au that present azimuthal extensions and form a double-ring structure. The 
comparison with radiative transfer models indicates that these two maxima of emission correspond to local increases in the 
dust surface density of about a factor 2.5 and 6.5 for the south and north clumps, respectively. The optically thick $^{13}$CO 
peak emission, which traces the temperature, and the dust continuum emission, which probes the disk midplane, additionally 
reveal two spirals previously detected in near-IR at the disk surface. The spirals seen in the dust continuum emission present, 
however, a slight shift of a few au towards larger radii and one of the spirals crosses the south dust clump. Finally, 
we present different scenarios in order to explain the complex structure of the disk.

\end{abstract}


\section{Introduction}

Planets form from the gas and solids contained in disks surrounding young stars. Once enough mass is assembled into 
a protoplanet, the gravitational interaction between this object and material in the disk is expected to generate a 
variety of substructures in the disk itself, as rings, spirals,  or cavities \citep{Kley2012, Baru2014, DiPi2016}. The 
formation of these substructures might in turn dramatically affect the formation of other planets through the creation 
of gas pressure maxima able to concentrate dust particles \citep{Pini2015, vdM2015, Guil2017}.

Observations of disks in nearby star forming regions at millimeter wavelengths and at high angular resolution have the 
capabilities to reveal these substructures across all the disk vertical height. ALMA is currently the most suited telescope 
for such observations and has unveiled azimuthal asymmetries \citep{vdM2013, Pere2014, Boeh2017}, rings \citep{ALMA2015, 
Isel2016, Fede2017}, cavities \citep{vdM2016a, Dong2017}, and spirals \citep{Chri2014,Tang2017, Pere2016}, in the dust 
and gas distribution.

A large variety of mechanisms has also been invoked to produce disc substructures without assuming the 
presence of protoplanets. Gaps/Rings might be zonal-flows \citep{Joha2009}, self-induced dust pile-ups due to radial migration 
\citep{Gonz2015}, pebble growth around condensation front \citep{Zhan2015}, or gradient in the viscosity at the outer edge 
of a dead-zone \citep{Rega2012,Ruge2016}. Local dust concentrations might be vortices induced by a large range 
of instabilities such the baroclinic instability \citep{Lesu2010} or the vertical shear instability \citep{Nels2013}, which do not 
require a planet. Importantly, in the case of a cavity + a horseshoe, the horseshoe might not be a vortex but might be induced 
naturally if the disc surrounds a binary system \citep{Ragu2017}. Finally, spirals might also be produced by the development of 
gravitational instabilities \citep{Krat2016}, by an external high-mass perturber undergoing a flyby \citep{Quil2005} or induced 
dynamically as a consequence of a different irradiation from the star due to a warp in the inner disc \citep{Mont2016}.

Due to the large range of scenarios able to produce substructures, observations at high angular resolution, 
using different tracers, are primordial. In this work, we show ALMA observations of the protoplanetary disk around MWC~758, that 
we also compare with previous images obtained in near-IR scattered light \citep{Beni2015}. The central star of the system MWC~758 
is an Herbig Ae star located at a distance of 151 $\pm$ 9 pc, based on the stellar parallax measured by the Gaia space telescope 
\citep[][]{Gaiaa2016,Gaiab2016}. This is less than the previous distance of 279$_{-58}^{+94}$ pc estimated from HIPPARCOS \citep{vanL2007}. 
The star is surrounded by an accretion disk characterized by a complex morphology. Despite showing strong infrared excess and an 
accretion rate of 10$^{-8}$ M$_{\odot}$ yr$^{-1}$, the disk has a large millimeter cavity of a few tens of au in radius 
\citep{Isel2010, Mari2015b}. This suggests that the cavity might have been generated by tidal interactions with companions 
and not photoevaporation by the central star \citep{Andr2011, Owen2011, Regg2017}. Furthermore, previous observations at 
radio-wavelengths have revealed the presence of asymmetries in the dust continuum emission possibly related to the accumulation 
of submillimeter grains in gas pressure maxima \citep{Mari2015b}. The disk also presents two spirals arms detected in 
near-IR scattered light \citep{Grad2013,Beni2015}, which might be interpreted as spiral density waves launched by 
planet(s) of a few $M_{Jup}$ \citep{Dong2015}.

Our new observations image the MWC~758 system in both the dust continuum emission, at a wavelength of $\sim$ 
0.88 millimeter, and in the $^{13}$CO and C$^{18}$O J = 3-2 emission lines. The observations achieve a resolution of 
0.1\arcsec-0.2\arcsec, or 15-30 au according to the distance of the system, which corresponds to a factor 4 of improvement 
with respect to previous millimeter-wave observations, and reveal unprecedented details of the disk morphology. In section 
2, we describe the acquisition of the data and the related reduction procedure, while in section 3 we present a morphological analysis 
of the maps of both the continuum and the molecular emission lines. In section 4, we constrain the physical properties of the 
circumstellar dust and gas by comparing the observations with radiative transfer models for the disk emission. The discussion of 
the results and the conclusion follow in section 5 and 6, respectively.


\section{Observations and data reduction}
\label{sec:obs}

We observed MWC 758 using ALMA in Cycle 3 at Band 7 as part of the project 2012.1.00725.S. Observations 
were performed on September 1 and 24, 2015, when 35 and 34 antennas were available, respectively. Baseline 
lengths ranged between 15.1 m and 1.6 km on September 1, and between 43.3 m and 2.3 km on September 24. The 
total on-source time spent on MWC 758 was about 58 minutes. 

The ALMA correlator was configured to record dual polarization with four separate spectral windows centered at 329.34, 
330.60, 341.01 and 343.01 GHz. The two first spectral windows were chosen to cover the molecular emission of C$^{18}$O 
J = 3 - 2 and $^{13}$CO J = 3 - 2, with a bandwidth for each spectral windows of about 234.4 MHz. The channel 
width in these two spectral windows is 61.0 kHz (0.055 km s$^{-1}$). The spectral resolution is twice the channel 
spacing since the data are Hanning smoothed. The third and fourth spectral window cover the continuum emission centered 
a 342 GHz ($\lambda$ $\sim$ 0.88 mm) with a total bandwidth of about 3.6 GHz.

The data were calibrated by the National Radio Astronomy Observatory (NRAO) staff using the CASA software package version 4.3.1 
\citep{Mcmu2007}. Simultaneous observations of the 183 GHz water line with the water vapour radiometers were used to reduce 
atmospheric phase noise before using J0510$+$180 and J0521$+$2112 for standard complex calibration. J0510$+$180 and J0423$-$0120 were 
used to calibrate the frequency-dependent bandpass. The flux scale was determined with observations of J0510$+$180, for which a flux 
of $\approx 2.3$ Jy was adopted. We self-calibrated the dust emission and then applied the results to the line emission. We used 
a time interval of 30 seconds for self-calibrating the phase and in a second step the scan length ($\sim$ 5 mn) for the amplitude 
calibration. 

For the final imaging steps, we used different weighting parameters on the visibilities to obtain the best rendering in term 
of sensitivity and angular resolution for each observations. We imaged the gas emission with natural weighting. This gives to each 
baseline a weight inversely proportional to its associated noise, privileging the short baselines. This yields an FWHM beam size of 
0.23\arcsec$\times$0.18\arcsec\ and an rms noise of 8 mJy beam$^{-1}$ per channel in $^{13}$CO and 11.0 mJy 
beam$^{-1}$ per channel in C$^{18}$O. Due to its higher signal-to-noise, the dust emission is imaged with modes giving more weight 
to longer baselines. First, we used a Briggs parameter equal to 0, an intermediate value, yielding an FWHM beam size of 
0.16\arcsec$\times$0.14\arcsec\ and an rms noise of 83 $\mu$Jy beam$^{-1}$. We then used the super-uniform 
mode, with a weight proportional to the baselines length, to better detect substructures. This gives an FWHM beam size of 
0.12\arcsec$\times$0.11\arcsec\ and an rms noise of 115 $\mu$Jy beam$^{-1}$.

\section{Results}
\label{sec:res}

\subsection{Dust continuum emission}
\label{sec:Dust}

\begin{figure*}
  \includegraphics[angle=-90,width=\textwidth]{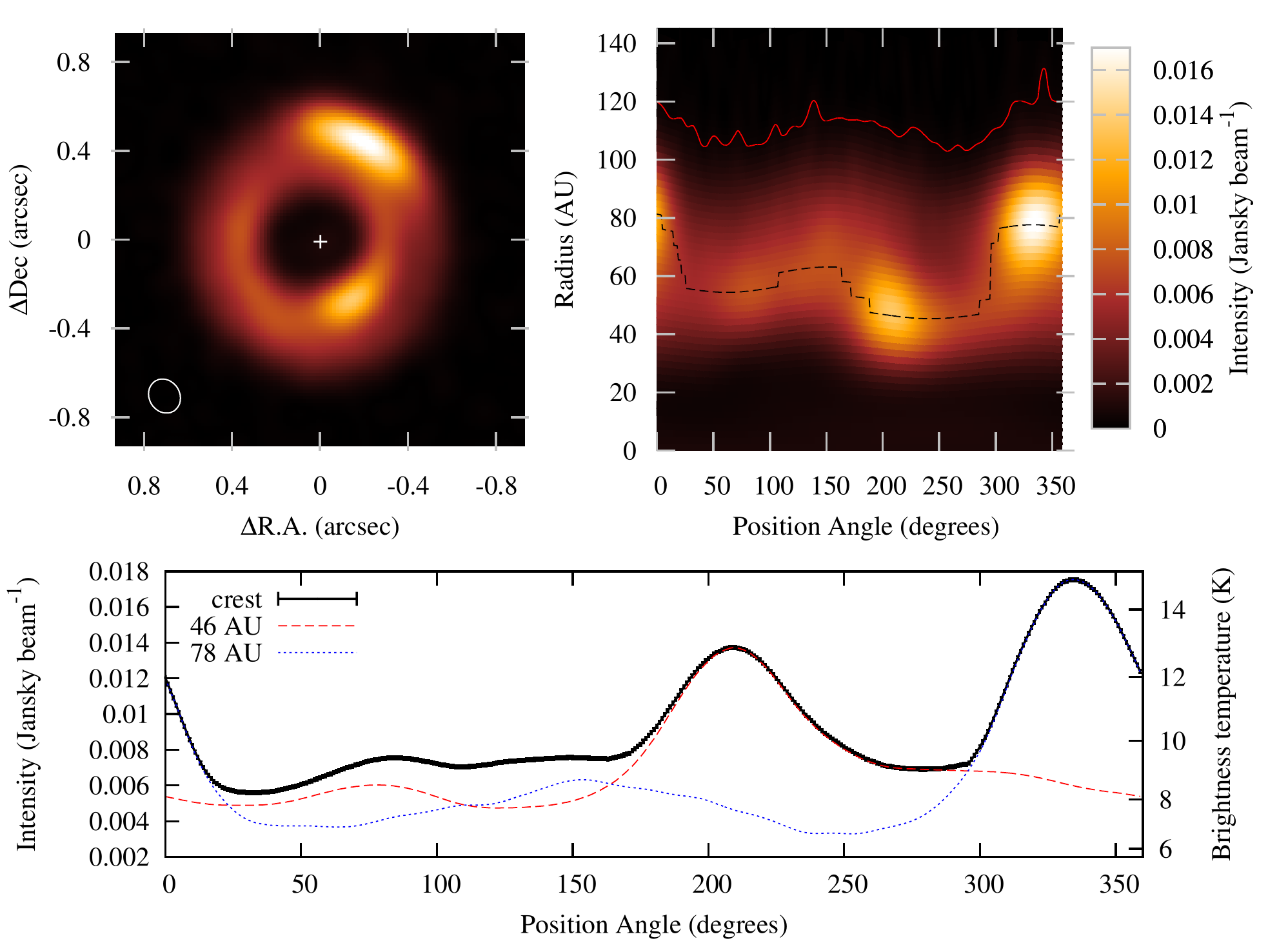}\\
  \caption{Top left: Map of the dust continuum emission recorded at a frequency of 342 GHz. The FWHM of the synthesized beam 
  is 0.16\arcsec$\times$0.14\arcsec (P.A. = 31.1$^{\circ}$) and is indicated by the white ellipse. Top right: Dust emission as 
  a function of the orbital 
  radius and of the position angle (P.A.) measured from the north to the east.  The red solid line is the 5 $\sigma$ contour, 
  corresponding to 0.4 mJy beam$^{-1}$, and the black dashed line indicates the crest of the emission, defined as the line that 
  connects the peaks of the emission measured at each P.A.. Bottom: Dust emission as a function of the P.A. The black solid line 
  is the emission along the crest. The dashed red and blue lines are at the radii of the south and north dust clumps, respectively.}
  \label{Fig:continuum}
\end{figure*}

The image of the dust continuum emission is shown in Figure~\ref{Fig:continuum} and was obtained adopting a Briggs weight parameter 
equal to 0. The white cross pinpoints the center of rotation as measured from the $^{13}$CO emission (see section \ref{sec:CO}) and 
is located at R.A. = 05$^h$30$^m$27$^s$.536 and decl. = +25$^\circ$ 19\arcmin\ 56\arcsec.70, cospatial with the Herbig star position. 
In the same figure, the top right panel displays the deprojected 
intensity as a function of the distance from the disk center and the position angle (P.A.), defined as the angle east of north 
(i.e towards the left). The orbital radius has been calculated by assuming an inclination of 21$^\circ$ and a P.A. of 62$^\circ$ 
for the disk. The black dashed line indicates the crest of the dust emission, which is defined as the line that connects the radial 
maximum of the emission for each position angle. 

The dust emission is characterized by a ring structure extending between roughly 40 and 100 au and two prominent azimuthal 
clumps. The brightest clump has a peak intensity of 17.2 mJy beam$^{-1}$ and is located in the north-west of the disk, at R = 81 au 
and P.A. = 335$^\circ$. The second clump has a peak intensity of 13.7 mJy beam$^{-1}$ and is located on the south-west, at a radius 
of 46 au and P.A. = 209$^\circ$. As shown by the deprojected flux map, the dust clumps correspond to the innermost and 
outermost radial positions of the dust crest, whose average radial distance from the disk center is 60 au. The brightness 
temperature of the north and south clumps, obtained by using the inverse of the Planck function, are 14.7 K and 12.9 K, respectively. 
These temperatures are lower, by a factor $\sim$ 4, than the predicted physical temperatures of the dust at the location of the 
clumps (see section \ref{sec:mod}), therefore suggesting that the dust emission is optically thin.

The bottom panel of Figure~\ref{Fig:continuum} indicates the intensity along the crest of the dust emission as well as the 
azimuthal profile of the dust emission at the radii of the two clumps. The contrast of the north and south dust clumps 
relative to the average emission at the same radial position is about 4 and 2.5, respectively.

\begin{figure*}
  \includegraphics[angle=-90,width=\textwidth]{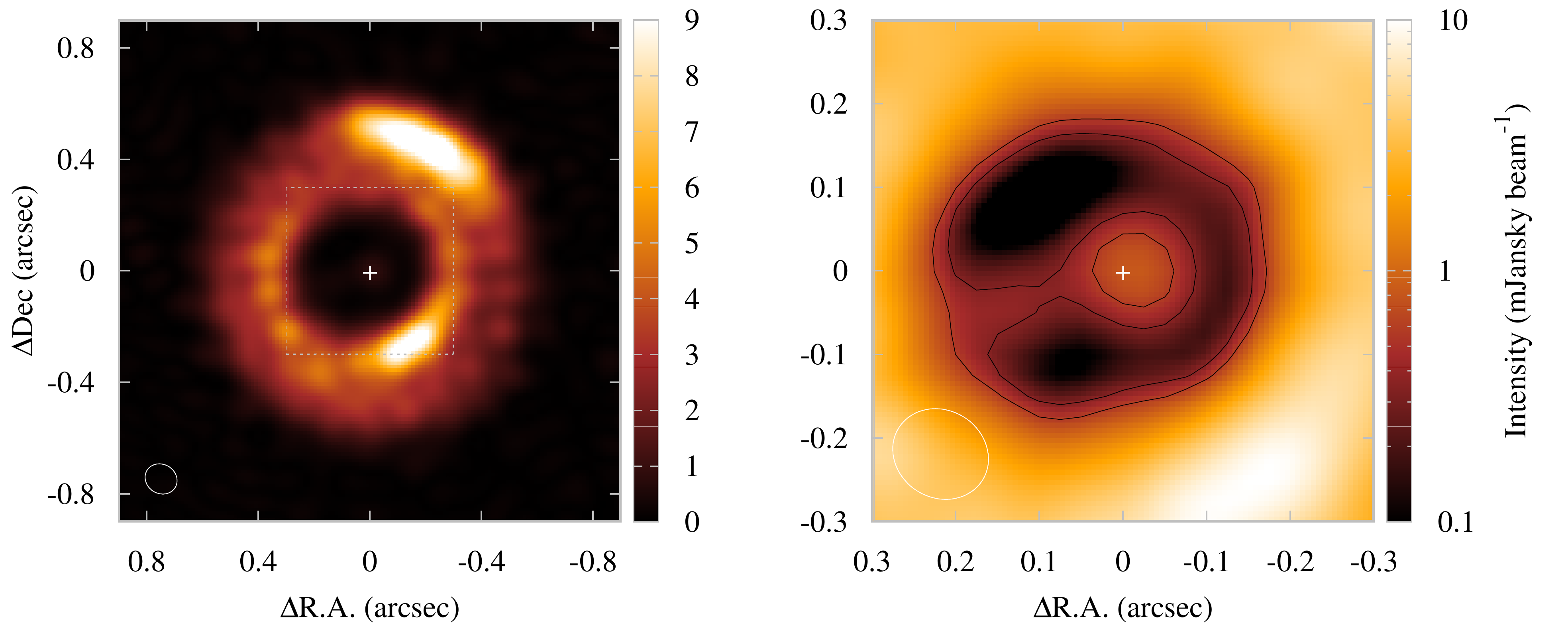}\\
  \caption{Left: Map of the dust continuum emission obtained using super-uniform weighting. The FWHM of the synthesized
  beam is 0.119\arcsec$\times$0.105\arcsec (P.A. = 62.2$^{\circ}$) and the noise is 115 $\mu$Jy beam$^{-1}$. The rectangle 
  in grey dashed line shows 
  the inner region displayed in the right panel. Right: Map of the disk cavity in logarithmic scale. The black solid lines 
  indicate the 3 and 5 $\sigma$ intensity contours. The white cross indicates the position of the center of rotation of the 
  system, derived from the $^{13}$CO J = 3-2 emission line.}
  \label{Fig:cavity}
\end{figure*}

Figure~\ref{Fig:cavity} shows the continuum map obtained with super-uniform weighting, which delivers a smaller beam size 
(0.12\arcsec$\times$0.11\arcsec) but a higher noise of 115 $\mu$Jy beam$^{-1}$, visible by dot-like features in the image. 
This new image better reveals the azimuthal morphology of the dust clumps. The peak emission is 12.3 mJy beam$^{-1}$ and 
10.2 mJy beam$^{-1}$ for the north and south clump, respectively. Converted in brightness temperature, the values are, 17.2 K 
and 15.3 K, slightly larger than with the previous weighting parameter Briggs = 0, indicating a better resolution, mainly 
radially, of the dust clumps structure. Also, the dust clumps appear to be linked to a double-ring structure, located at the same 
radii. Finally, the continuum emission presents also substructures that may trace faint spirals in the dust. A detailed discussion 
of these features and a comparison with the $^{13}$CO peak emission line and previous IR observations are done in section 
\ref{sec:scattering}. 

On the right panel, an inset toward the disk center shows a non-resolved source of continuum emission detected inside the cavity. 
The peak emission of the signal is 0.83 mJy beam$^{-1}$, about 7.2 times the noise level in the outer regions of the map. Due 
to the close proximity with the center of rotation of the disk, this emission probably comes from an inner disk in rotation 
around the Herbig Ae star. This is consistent with the spectral energy distribution of the disk that present near-IR emission, 
and traces warm dust close to the star \citep{Eisn2004, Isel2008}. As the signal is not resolved, this implies an outer radius 
$\leq$ 8 au for the inner disk.

\subsection{$^{13}$CO and C$^{18}$O J=3-2 emission}
\label{sec:CO}

\begin{figure*}
  \includegraphics[angle=-90,width=\textwidth]{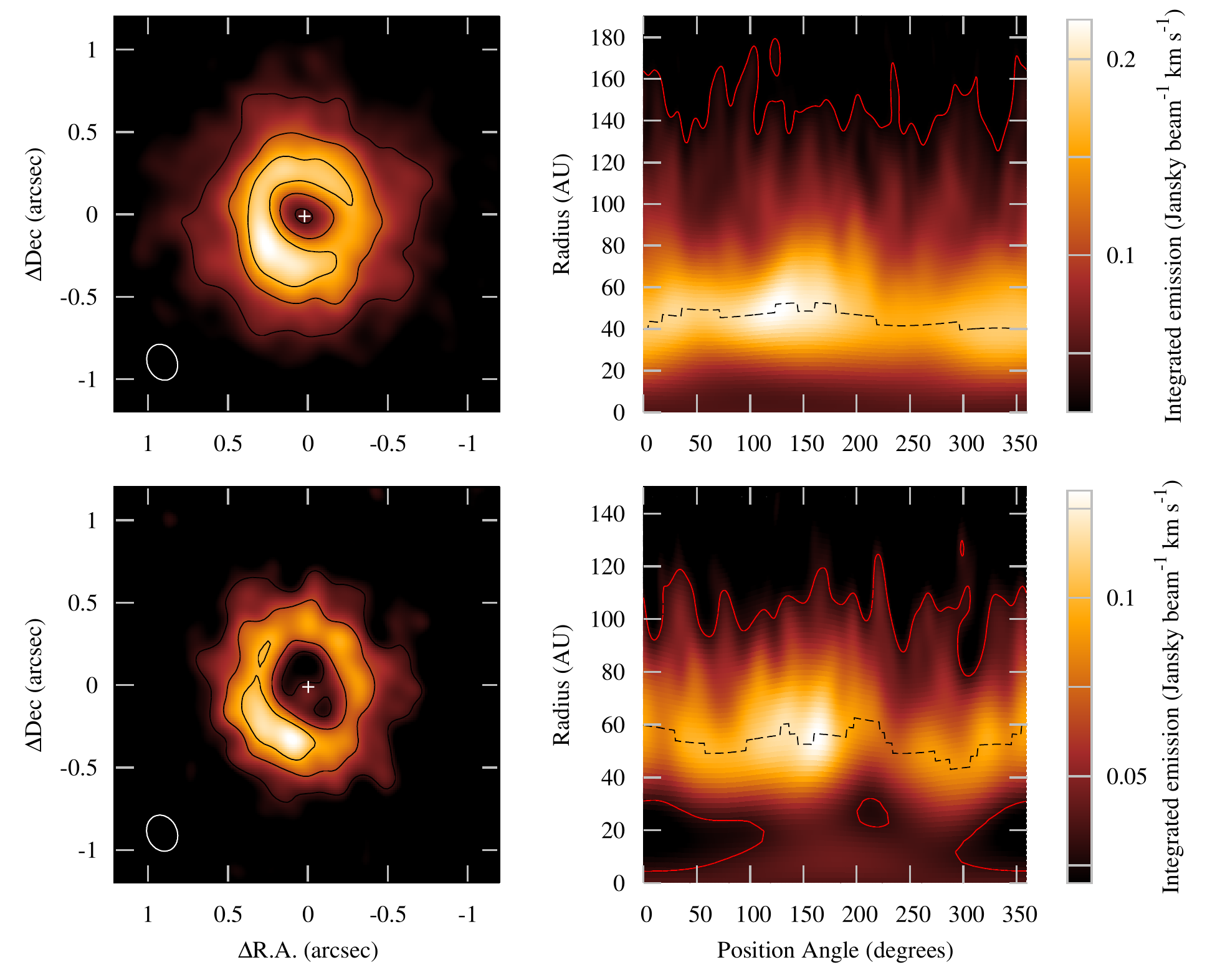}\\
  \caption{Top left: $^{13}$CO J=3-2 integrated intensity (Moment 0) map. The FWHM of the synthesized beam is 
  0.23\arcsec$\times$0.18\arcsec\ (P.A. = 26.2$^{\circ}$) and is indicated by the white ellipse. The contours correspond to 
  25, 50 and 75 $\%$ of the peak emission. Top right: $^{13}$CO J=3-2 integrated intensity in function of radius and P.A.. The 
  dashed line represents the $^{13}$CO crest position in the disk. The red line indicates the 3 $\sigma$ contours with $\sigma$ 
  $=$ 8.2 mJy beam$^{-1}$ km s$^{-1}$ for $^{13}$CO and 9.8 mJy beam$^{-1}$ km s$^{-1}$ for C$^{18}$O. Bottom 
  panels: Same as on the top but for the C$^{18}$O J=3-2 emission line.} 
  \label{Fig:mom0CO}
\end{figure*}

The $^{13}$CO and C$^{18}$O emission lines were imaged using a Briggs factor of 2, to optimize the signal-to-noise ratio. 
The top and bottom left panels in Figure~\ref{Fig:mom0CO} show the moment 0 maps after continuum subtraction. The white 
cross represents the center of rotation of the disk, as derived from the $^{13}$CO velocity map (Figure~\ref{Fig:mom12} in 
section~\ref{sec:vel}). The $^{13}$CO emission is more radially extended than the dust continuum, with a flux detected up 
to about 140 au from the star. On the contrary, the radial extent of the C$^{18}$O emission is about 100 au, similarly to 
the dust. This difference is likely due to the lower optical depth of C$^{18}$O compared to $^{13}$CO.

The right panels in Figure~\ref{Fig:mom0CO} present the emission, the crest, and the 3 $\sigma$ contours as a function 
of the position angle. The radius of the crest varies between 40 and 50 au for $^{13}$CO and from 50 to 60 au for C$^{18}$O. 
This suggests that the gas surface density increases from the disk center to about 50-60 au. The crest radius 
for the two molecules is almost constant as a function of the azimuth, with variations no larger than 10 au. In particular, 
we do not observe any increase in the CO emission at the position of the two dust clumps. 

The velocity map of $^{13}$CO is shown in Figure~\ref{Fig:mom12}, in section \ref{sec:vel}, and indicates a position 
angle of $\sim$ 62 $\pm$ 2$^{\circ}$. If the IR features observed in \cite{Grad2013} and \cite{Beni2015} are trailing spirals, 
the disk is rotating clockwise and the north-west 
part of the disk is the closest to us. This seems confirmed by the position angle at which the emission of the two molecules 
reaches a maximum, between roughly 90$^{\circ}$ and 180$^{\circ}$, with a value of 223 mJy beam$^{-1}$ km s$^{-1}$ at PA = 
125\arcdeg\ for $^{13}$CO, and 132 mJy beam$^{-1}$ km s$^{-1}$ at PA = 160\arcdeg\ for C$^{18}$O. If the south-east side is 
indeed the further half, the inner wall of the disk in this direction is directly exposed to the observer. The azimuthal 
minimum in gas is at a P.A. of about 240\arcdeg, with a value 1.4 times lower than the maximum in $^{13}$CO and 1.9 times 
the maximum in C$^{18}$O. 

\begin{figure*}
  \includegraphics[angle=-90,width=0.95\textwidth]{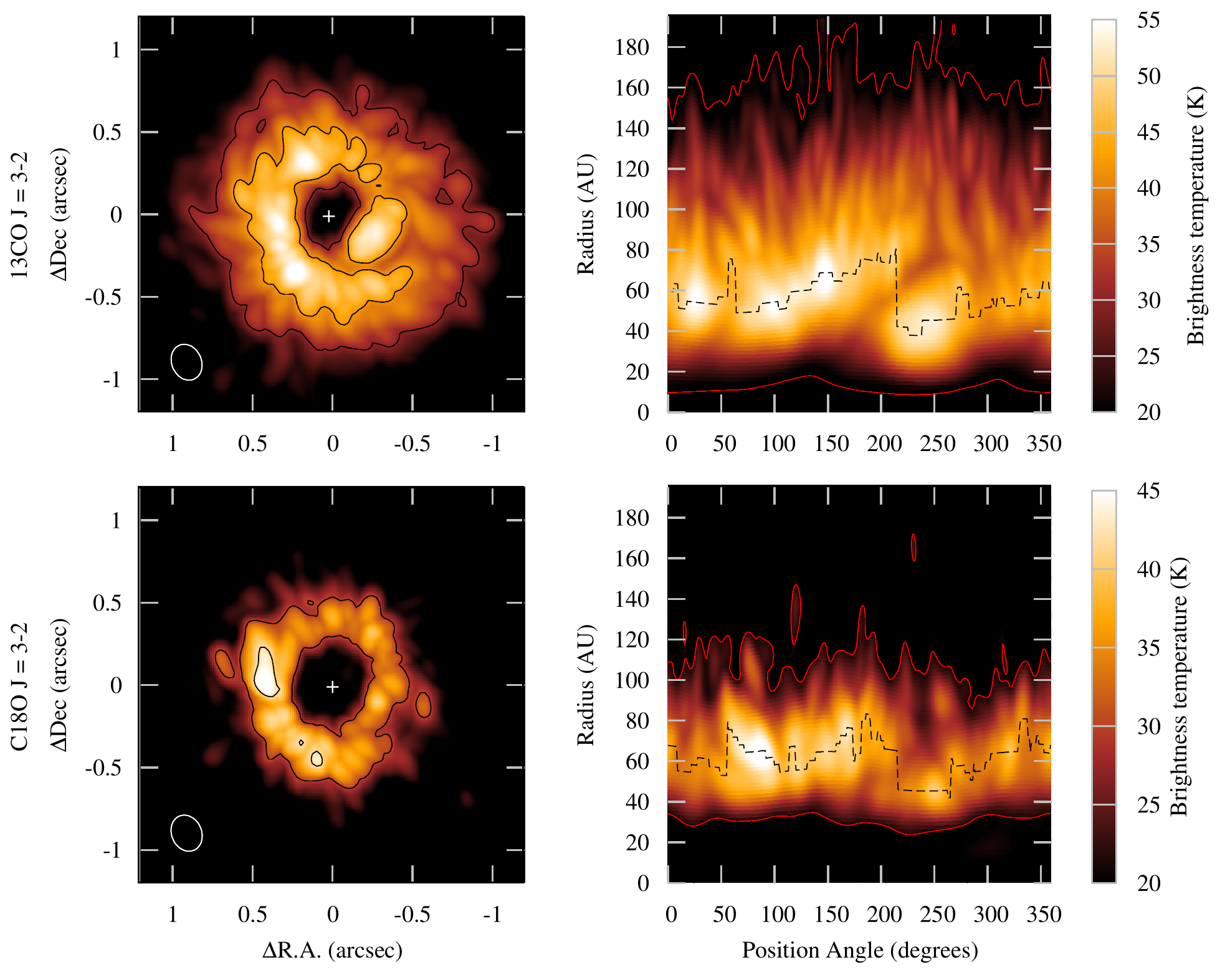}\\
  \caption{Left: maps of the $^{13}$CO and C$^{18}$O J = 3-2 peak emission line in units of the brightness temperature. The contours 
  are plotted at 30 K and 42 K. Right: Peak emission line as a function of the azimuth and of the radial distance. 
  The red line indicates the 3 $\sigma$ contours with $\sigma$ $=$ 8.2 mJy beam$^{-1}$ for $^{13}$CO and 10.2 mJy 
  beam$^{-1}$ for C$^{18}$O, or 17.3 K and 19.0 K, respectively, by using the inverse of the Planck function.}
  \label{Fig:mom8}
\end{figure*}

In Figure~\ref{Fig:mom8}, we show the peak intensity maps of the $^{13}$CO and C$^{18}$O emission lines. The $^{13}$CO peak emission 
is our most optically thick tracer and reveals a large spiral arm, as indicated by the increasing radius of the crest of the emission 
as a function of the position angle, between P.A. = 50$^{\circ}$ and 200$^{\circ}$. A secondary spiral arm may also be visible at 
P.A. = 220-360$^{\circ}$ but with less contrast with the surrounding disk. Along the crest, the brightness temperature of the emission 
varies between 40-60 K but shows little dependence on the orbital radius. In section \ref{sec:mod}, we will show that these values of 
the brightness temperature are consistent with optically thick emission, with $\tau$ $\sim$ 3.7, and therefore indicate variations in 
the temperature. 

In C$^{18}$O, the observations present similar but less pronounced spirals. The crest of the emission, indicated by the black dashed 
line, is more centered at the position of 60 au, as for the integrated emission. This is probably due to the lower optical depth of this 
tracer, with a maximum optical depth of $\sim$ 0.6 according to our model in section \ref{sec:mod}. It is then less sensitive to 
variation in the temperature than $^{13}$CO and better traces the gas surface density. This is similar to the integrated emission 
maps, which are more optically thin than the peak emission maps, and do not present spirals.

As a note, the peak emission map was obtained without subtracting the continuum emission, that would have been resulted in underestimating 
the line peak brightness temperature (see the discussion in \citealt{Boeh2017}). However, the differences between the two methods 
are minor in MWC~758 due to the low optical depth of the dust. 

\subsection{Disk dynamic: an inner warped disk?}
\label{sec:vel}

\begin{figure*}
  \includegraphics[angle=-90,width=0.95\textwidth]{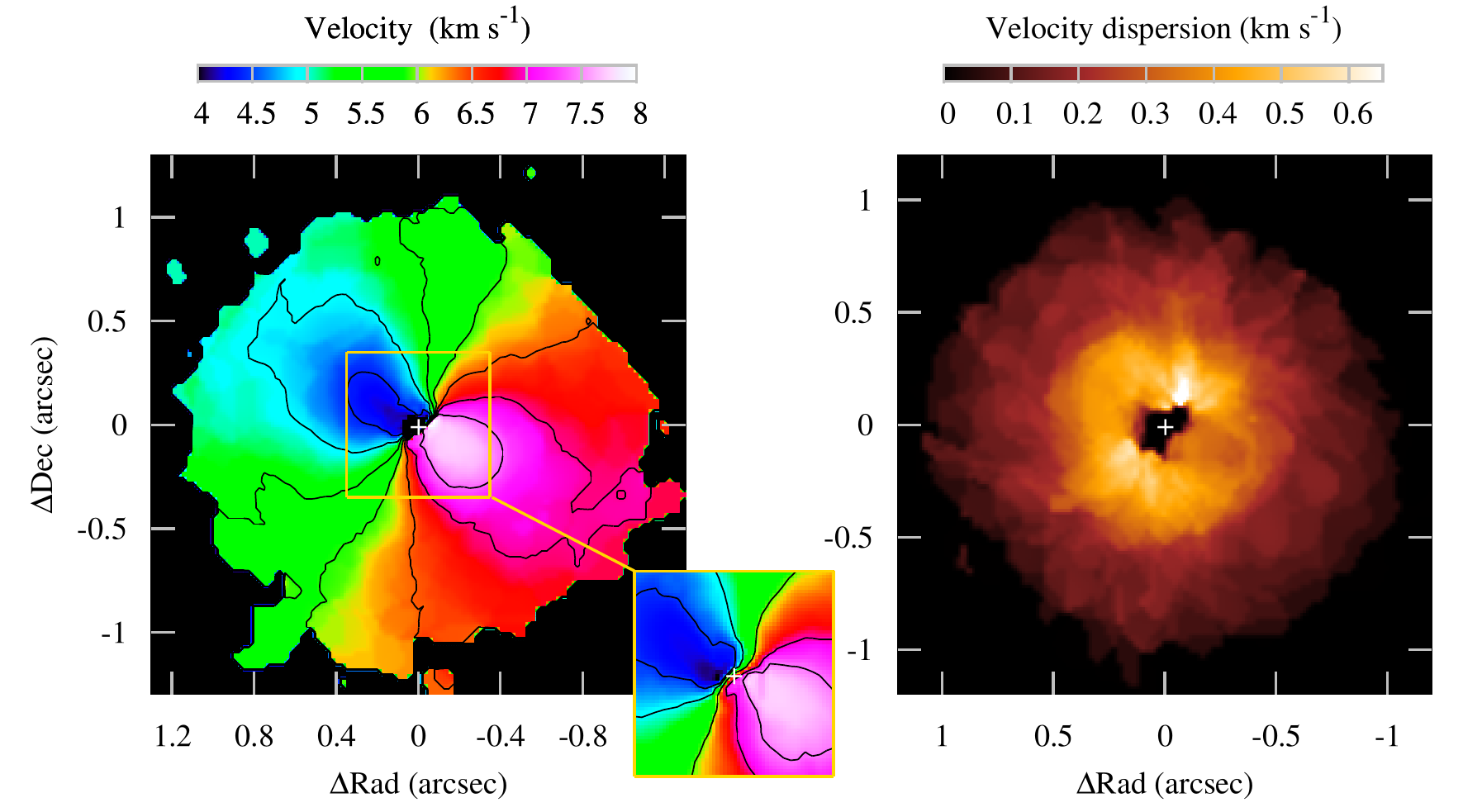}\\
  \caption{Left: Map of the velocity centroid for the $^{13}$CO J = 3 - 2 line plotted between 4 to 8 km s$^{-1}$. The systemic velocity 
  is 5.90 km s$^{-1}$ and the contours are displayed from 4.40 km s$^{-1}$ to 7.40 km s$^{-1}$ by paths of 0.5 km s$^{-1}$. The inset shows 
  the inner regions of the velocity map, taking into account all the channels above 3 $\sigma$ (with $\sigma$ = 8 mJy beam$^{-1}$)
  instead of 5 $\sigma$ for the larger map. Right: map of the velocity dispersion in $^{13}$CO plotted between 0 to 0.65 km s$^{-1}$.}
  \label{Fig:mom12}
\end{figure*}

We present in Figure \ref{Fig:mom12} the kinematic of the disk. The major axis of the disk has a 
position angle of 62 $\pm$ 2 degrees and the disk has a systemic velocity of 5.90 $\pm$ 0.05 km s$^{-1}$.
The velocity map is globally consistent with a disk in Keplerian rotation around a central star of 1.4 $\pm$ 
0.3 solar masses, considering an inclination of 21 degrees \citep{Isel2010}.  Inside the dust cavity, 
however, the gas presents hints of departure from Keplerian velocity. The twist of the iso-velocity curves might 
reveal radial inflow of the gas or a warp linking the outer disk with a non-coplanar inner disk, where a 
non-resolved source of dust emission is also observed.

The panel in the right side of Figure \ref{Fig:mom12} shows the $^{13}$CO velocity dispersion. The velocity dispersion 
usually decreases with the radius, principally due to lower temperatures. It can also vary azimuthally, with a larger range 
of projected velocities probed along the minor axis. We nevertheless also observe a minimum along the west 
side of the major axis, with values of 0.3 km s$^{-1}$ compared to 0.4 km s$^{-1}$ along the east side. We previously 
pointed out in the integrated $^{13}$CO and C$^{18}$O emission maps an azimuthal minimum along this direction. Such diminution 
is consistent with an azimuthal dip of the temperature. However, it appears at a first glance surprising that the $^{13}$CO peak 
emission, which is our most optically thick tracer, shows a bright emission along this azimuthal direction, at $\sim$ 40 au, at 
the inner edge of the outer ring. This suggests that the azimuthal decrease in emission might be due to a shadow effect from the 
inner edge of the outer disk, where a spiral is present and might have a higher scale height, 
or from an inner warped disk.  

To estimate qualitatively the hypothesis of an inner warped disk, we compared our observations with the 
simulations of \cite{Facc2018}. They performed synthetic images of disks having a relative inclination of 30$^\circ$ and 
70$^\circ$. The case with a moderate inclination predicts variations in the iso-velocity curves 
similar to what is observed in MWC~758 (i.e. their figure 17). Also, an inner disk with a mild relative inclination will 
project a large shadow toward one side of the disk, but still letting an inner central region to be irradiated by the 
central star due to the relative geometry of the two disks (see Fig. 9 and 11 in \citealt{Facc2018}). Our ALMA data are 
then consistent with an inner mildly warped disk whose west side is the closer side. This is also supported by previous 
near-IR interferometric observations of \cite{Eisn2004} and \cite{Isel2008} that estimated the inclination of the inner regions 
to be in the range 30${^\circ}$-40${^\circ}$, instead of 21${^\circ}$ $\pm$ 2 for the outer regions \citep{Isel2010}. Future CO 
emission lines observations 
with a better angular resolution and sensitivity, perhaps using higher rotational lines, will help to constrain the velocity 
structure in the cavity. This case can 
be compared to the disk HD~135344B which also probably has a mildly inner warped disk \citep{Stok2016}. It is different 
of the binary system HD~142527, which contains a strongly warped disk, by 70$^\circ$ \citep{Casa2015a}, and projects two shadows 
in the outer regions separated azimuthally by $\sim$ 180$^\circ$ \citep{Mari2015a, Casa2015b, Boeh2017}. \\

\section{Analysis}

\subsection{Modeling of the dust and gas emission}
\label{sec:mod}






\begin{figure}
  \includegraphics[angle=0,width=0.5\textwidth]{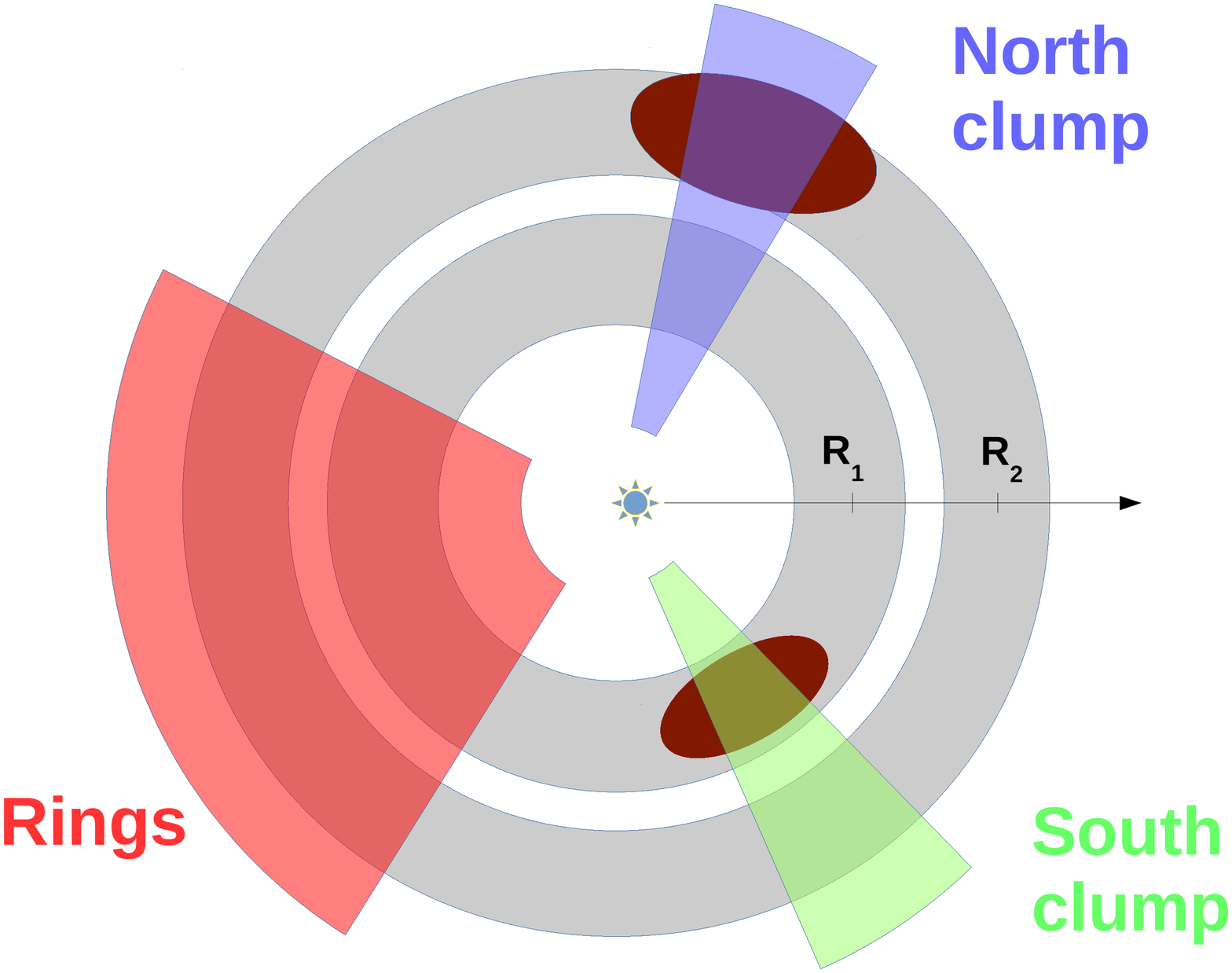}\\
  \caption{Sketch of the MWC~758 system with the three azimuthal directions used for the disk modeling. The underlying rings are 
  modeled using the emission in the angular range 75$^{\circ}$ - 150$^{\circ}$ represented by the red area. In green, the 
  model across the south dust clump in the angular range 200$^{\circ}$ and 220$^{\circ}$, and in blue the model across the north 
  dust clump in the angular range 325$^{\circ}$-345$^{\circ}$.} 
  \label{Fig:sk}
\end{figure}

The disk around MWC~758 has a complex morphology, characterized by dust clumps and spirals surrounding a large cavity. A 
complete modeling of the 3D structure of the disk would be very difficult. We performed instead a study which focuses 
on three particular azimuthal directions, as shown in Figure~\ref{Fig:sk}. The emission coming from the red area will 
be used to characterize the underlying rings, while we will study the variations in dust and gas surface densities along 
the north and south dust clumps, inside the blue and green areas, respectively. 

\begin{figure*}
  \includegraphics[angle=-90,width=\textwidth]{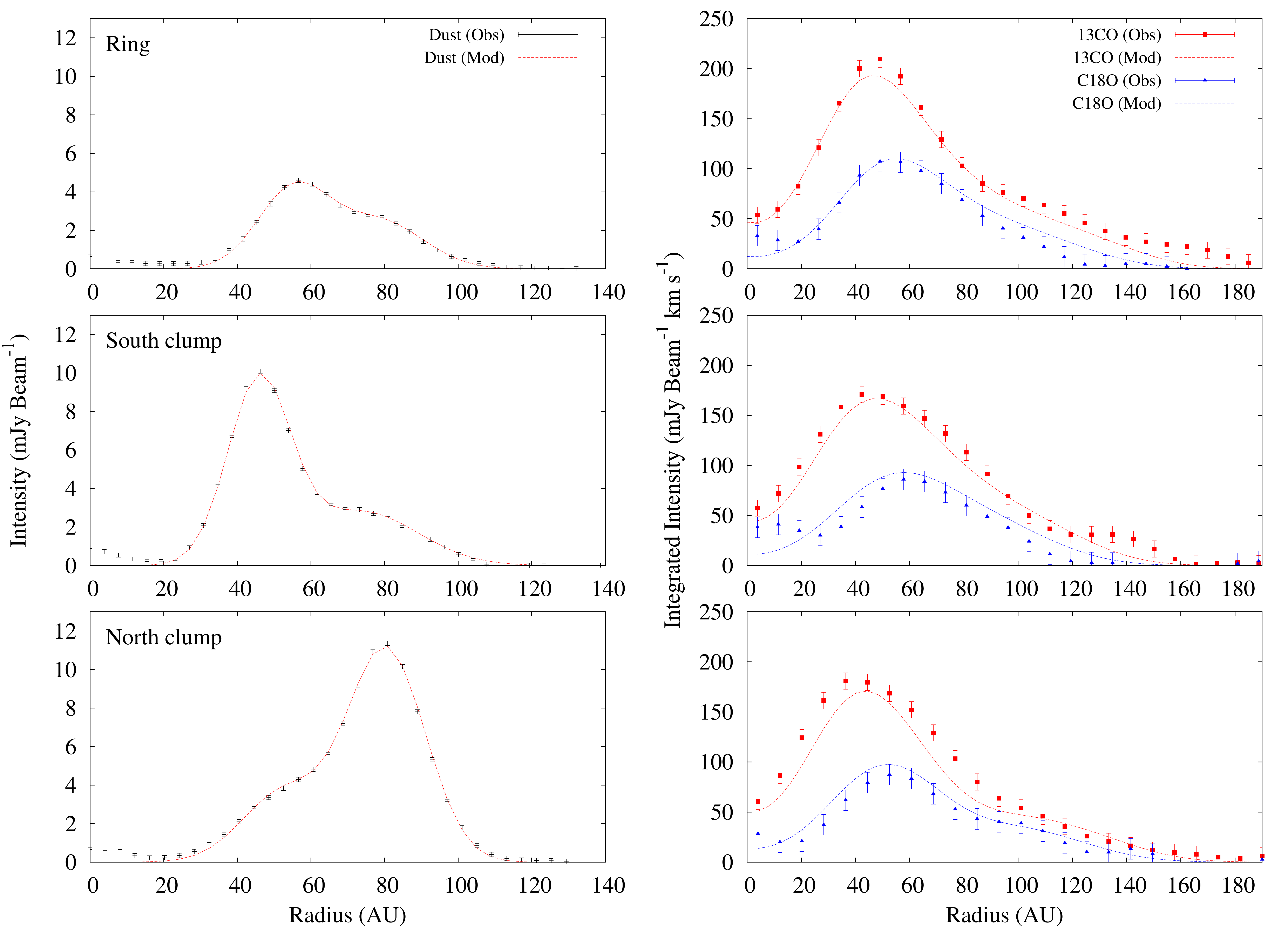}\\
  \caption{Left: Dust continuum emission obtained with super-uniform weighting in function of the radial distance (top, P.A. 
  = 75-150$^{\circ}$, Middle, P.A. = 200-220$^{\circ}$, Bottom, P.A. = 325-345$^{\circ}$). The observations and their associated 
  error bars are in black, while the models are represented by red dashed lines. Right: $^{13}$CO and C$^{18}$O J = 3-2 integrated 
  emission lines (Mom 0). The observations are represented by a red square for $^{13}$CO and a blue triangle for C$^{18}$O, with 
  their error bars, and the models are indicated by dashed lines.} 
  \label{Fig:int}
\end{figure*}

The model discussed below follows the analysis presented in \cite{Boeh2017}. Following the observations of the dust with 
super-uniform weighting, we assume a surface density structure for the dust and the gas represented by the sum of two Gaussians, 
using the formula:

\begin{align}
  \Sigma(r) & = \Sigma_{1}~\mathrm{exp}\left( -\left( \frac{r-R_1}{w_{1}} \right)^2 \right) \nonumber  \\
            & + \Sigma_{2}~\mathrm{exp}\left( -\left( \frac{r-R_2}{w_{2}} \right)^2 \right) 
\end{align}
with $R_1$ and $R_2$, the radial positions of the two Gaussians, $\Sigma_1$ and $\Sigma_2$ their surface density, and $w_1$ and 
$w_2$, their half-widths. The dust surface density is then described by using 6 free parameters. For the gas, less spatially 
resolved, we generally only use 4 free parameters, fixing the radial location of the two Gaussians at the same position as for 
the dust. However, in one case, along the south clump, we had to modify the radial position of the inner Gaussian on a less eccentric 
position to better represent the gas emission. Vertically, the gas is assumed in hydrostatic equilibrium, at the midplane temperature, 
and the dust vertically mixed with the gas. We also tested two different degrees of dust settling, with the dust vertical scale height 
reduced by a factor of 5 and 10, but only negligible changes were observed between the models.

\begin{center} 
 \begin{table}  
  \begin{tabular}{|c|cccccc|}
    \hline
          & $\Sigma_1$ & $R_1$  & $w_1$ & $\Sigma_2$ & $R_2$  & $w_2$ \\ 
          & $\mathrm{(cm^2 g^{-1})}$ & (au) & (au) &  $\mathrm{(cm^2 g^{-1})}$ & (au) & (au) \\ \hline
          \multicolumn{7}{|l|}{Ring (P.A. = 75$^{\circ}$-150$^{\circ}$)} \\ \hline
          Dust & 0.040  & 57 & 8  & 0.036  & 81 & 12 \\
          Gas  & 1.3    & 57 & 17 & 0.9    & 81 & 30 \\ \hline
          \multicolumn{7}{|l|}{South clump (P.A. = 200$^{\circ}$-220$^{\circ}$)} \\ \hline
          Dust & 0.102 & 47 & 5  & 0.030 & 78 & 18 \\
          Gas  & 0.9   & 53 & 17 & 0.9   & 78 & 24 \\ \hline      
          \multicolumn{7}{|l|}{North clump (P.A. = 325$^{\circ}$-345$^{\circ}$)} \\ \hline
          Dust & 0.027 & 56 & 12 & 0.24  & 82 & 9 \\
          Gas  & 1.3   & 56 & 18 & 1.0  &  82  & 28 \\ \hline    
  \end{tabular}
  \caption{Density prescription for the Gaussians describing the dust and gas surface density. }     
  \label{tab:dens}
\end{table}
\end{center} 

\begin{figure}
  \includegraphics[angle=270,width=0.48\textwidth]{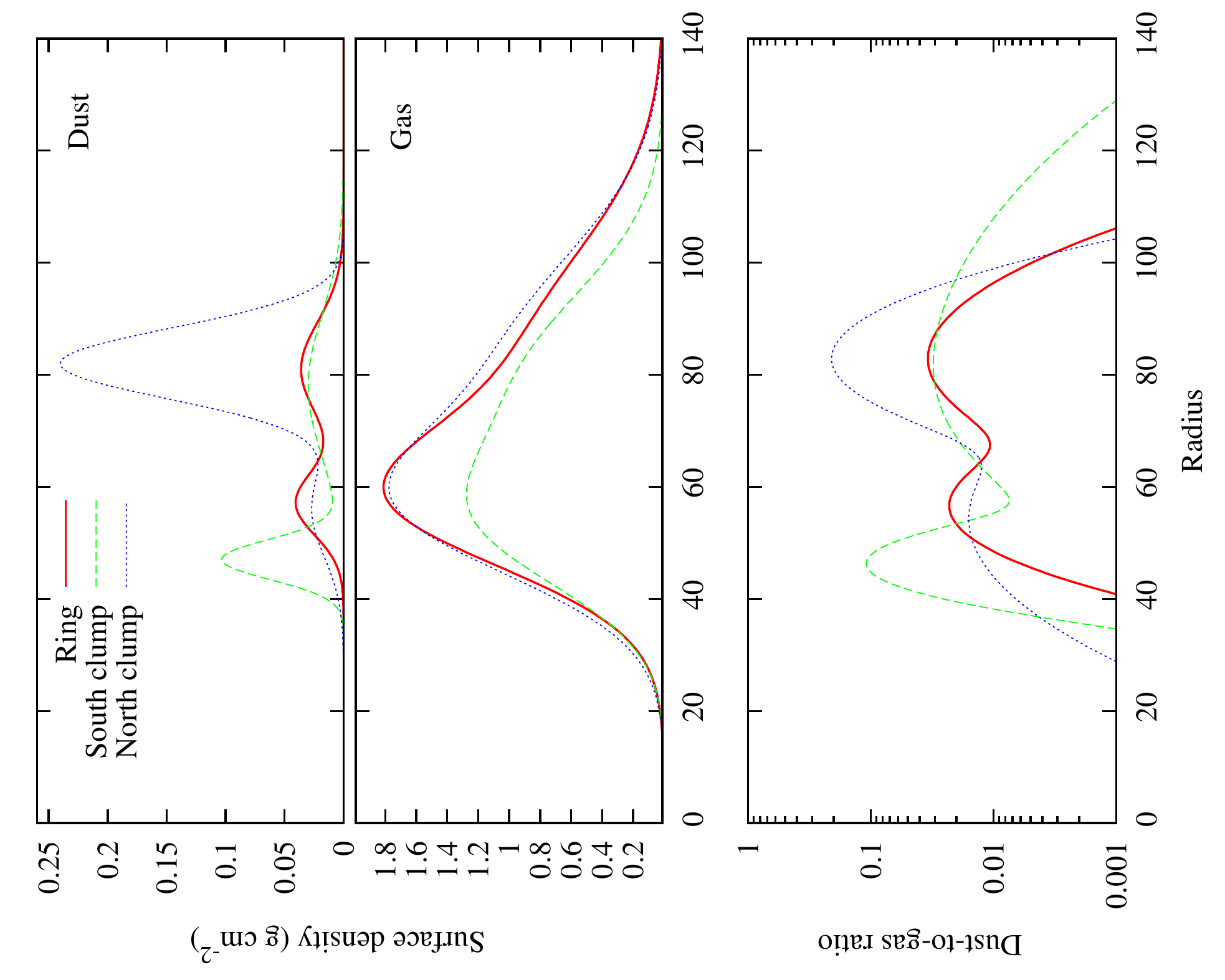}\\
  \caption{Top and middle panels: Dust and Gas surface densities across the rings (P.A. = 75-150$^{\circ}$), the south dust 
  concentration (P.A = 200-220 $^{\circ}$), and across the north dust concentration (P.A. = 325-345 $^{\circ}$). The 
  bottom panel indicates the dust-to gas ratio for the same three position angles.}
  \label{Fig:dens}
\end{figure}

The temperature in the disk is determined by carrying out a two-dimensional (r,z) radiative transfer using the code RADMC-3D 
\citep{Dull2012}. The central star is an Herbig A5e star, with a temperature at the surface of 8130 K \citep{vdA1998}. Taking 
into account the new estimated distance of 151 pc, the intrinsic luminosity is now 15.3 L$_{\odot}$ with a radius for the star 
of 2.0 R$_{\odot}$. The mass, deduced from Keplerian rotation, is 1.4 $\pm$ 0.3 M$_{\odot}$. The stellar emission is mainly 
absorbed in protoplanetary disks by $\mu$m-size grains, which are expected to be well coupled to the gas. We then assume for the radiative 
transfer the standard gas-to-dust ratio of 100, which corresponds to an opacity of about 6 cm$^{2}$ g$^{-1}$ of gas, in the frequency 
range 0.1 - 1$\mu$m, at which most of the stellar energy is emitted. The dust grains responsible of the dust emission at millimeter 
wavelength are represented by a grain size distribution following a power law of exponent -3.5 between 0.05 $\mu$m and 1 mm, and 
a chemical composition corresponding to the solar elemental abundance \citep{Ande1989}. These dust properties were previously used 
in \cite{Boeh2017} and yields an opacity at 0.88 millimeter of 2.9 cm$^{2}$ g$^{-1}$. The fractional abundances of $^{13}$CO and 
C$^{18}$O, with respect to molecular hydrogen, are assumed to be 9$\times$10$^{-7}$, and 1.35$\times$10$^{-7}$, respectively 
\citep{Qi2011}.

The radial profiles of the dust continuum emission are displayed in the left panels of Figure~\ref{Fig:int} and the integrated 
emission of the $^{13}$CO and C$^{18}$O J = 3-2 in the right panels, from top to bottom for each of the three sectors presented 
in Figure \ref{Fig:sk}. We use the integrated emission to fit the gas surface density distribution, as its optical depth is 
lower than for the peak emission, and less sensitive to variations at the disk surface, such than spirals, that are not taken 
into account in our modeling. At a first glance, we observe large azimuthal variations in the dust emission while the $^{13}$CO 
and C$^{18}$O integrated emissions are more constant with the position angle. 

The models that best fit the data are represented by dotted lines and were found manually. They generally well recover 
the disk emission. A summary of the model parameters that give the dust and gas surface densities are provided 
in Table~\ref{tab:dens} and in Figure~\ref{Fig:dens}. Figure~\ref{Fig:dens} also shows the radial evolution of the dust-to-gas 
ratio. On the left panels of Figure~\ref{Fig:int}, the 
dust emission, optically thin, presents a double-ring structure. The two rings are located at 57 au and 81 au and have a comparable 
surface density of $\sim$ 0.4 g cm$^{-2}$. In comparison, the two dust clumps are on more eccentric positions with the south clump 
at 47 au and the north clump at 82 au and have surface densities of about 2.5 and 6.5 times, respectively, the values encountered in 
the underlying rings. 

\begin{figure}
  \includegraphics[angle=-90,width=0.5\textwidth]{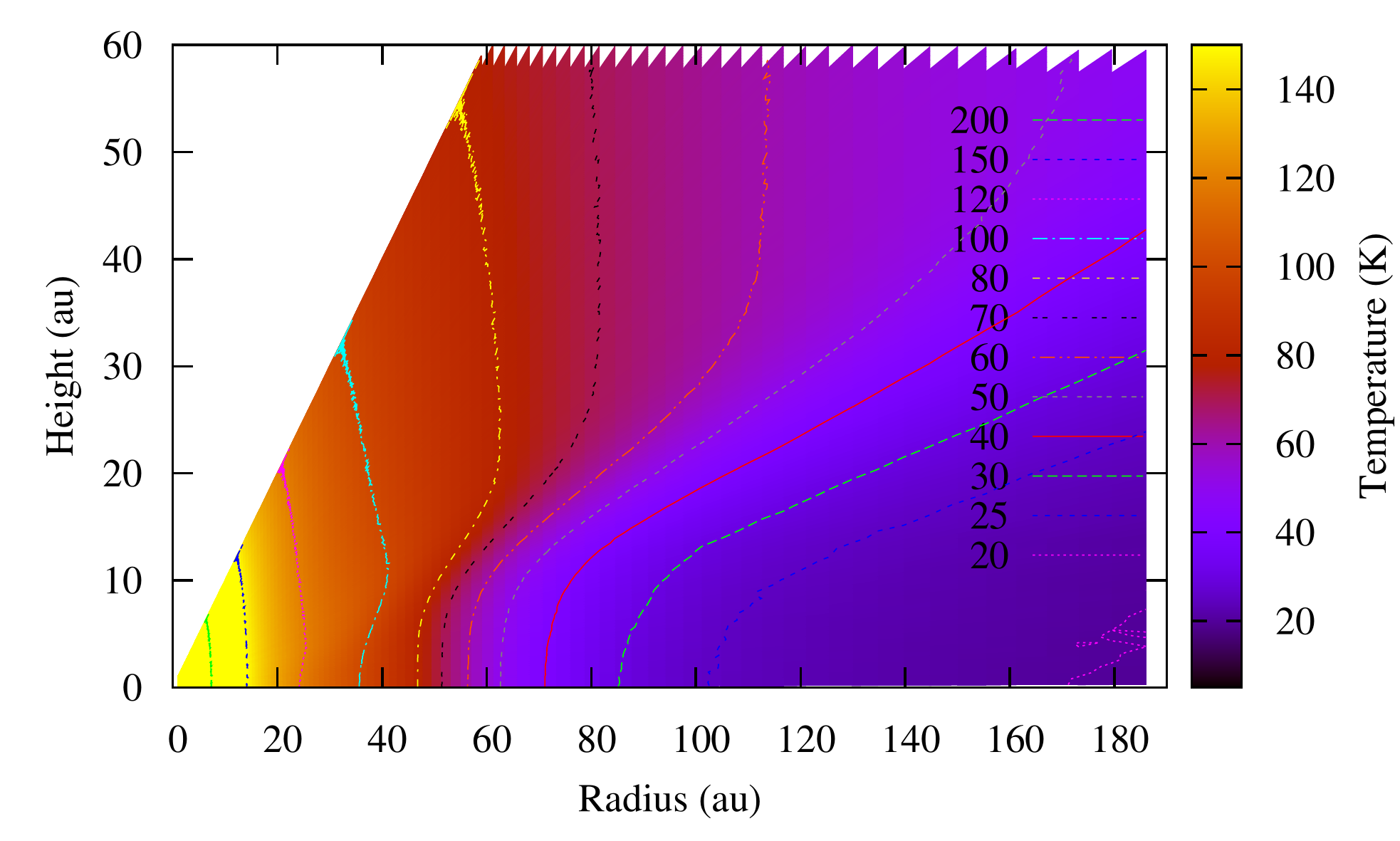}\\
  \caption{Temperature in the disk at P.A. between 75$^{\circ}$ and 150$^{\circ}$.} 
  \label{Fig:temp}
\end{figure} 

The gas has small azimuthal variations in the surface density structure. As we consider that the small grains follow the gas, 
the temperature of the disk is almost invariant azimuthally. The temperature along the double-ring structure is given 
in Figure~\ref{Fig:temp}. Only the gas surface density along the south dust clump appears slightly smaller. This 
area is located at 
the azimuthal minimum in the $^{13}$CO and C$^{18}$O integrated emission. As mentioned in section \ref{sec:vel}, this region might 
in fact be in the shadow projected by the inner regions of the disk. In this case, the temperature would be locally lower and 
inversely, the gas surface density higher, more similar to the value encountered at other polar angles.

The gas distribution is also more extended radially than the large dust grains distribution, both in the cavity and in the outward direction.
The lower resolution obtained for the gas of $\sim$ 0.2\arcsec\, about twice larger than for the dust, and its lower signal-to-noise,
does not allow to probe as efficiently substructures. Therefore, we are not able to observe if the double-ring structure also exists 
in the gas distribution. We also do not find evidence of gas concentration cospatial with the dust clumps. This leads to an 
increase of the dust-to-gas ratio in the dust clumps to a value of about 0.1, shown in the bottom panel of Fig. \ref{Fig:dens}, ten 
times the usual standard value. As a simple estimation of our precision, C$^{18}$O, our best tracer of the gas density, has maximum 
values of about 100 mJy beam$^{-1}$ km s$^{-1}$, about 10 times larger than the rms noise. The precision of the gas surface density 
at 3 $\sigma$ is then known at best at $\pm$ 30\%. 

Considering an average temperature of 60 K at a distance of 60 au, where the gas density reaches its maximum, we find optical 
depths of about $\sim$ 3.7 and 0.6 at the center of the $^{13}$CO and C$^{18}$O J=3-2 emission lines, respectively. By comparison, 
the dust has a maximum optical depth of about 0.12 in the underlying double disk structure, but which raises up to 0.7 in the north 
dust clump. Finally, if we consider that the gas and the dust distribution in the disk are represented azimuthally at 70\% by 
the red sector, and at 15\% each by the sectors along the dust clumps, the gas mass in the disk is about 1.3$\times$10$^{-3}$ 
M$_\odot$ and the dust mass about 3.0$\times$10$^{-5}$ M$_\odot$.


\subsection{Comparison of the dust and $^{13}$CO peak emission with the IR Scattered Light}
\label{sec:scattering}

\begin{figure*}
  \includegraphics[angle=-90,width=\textwidth]{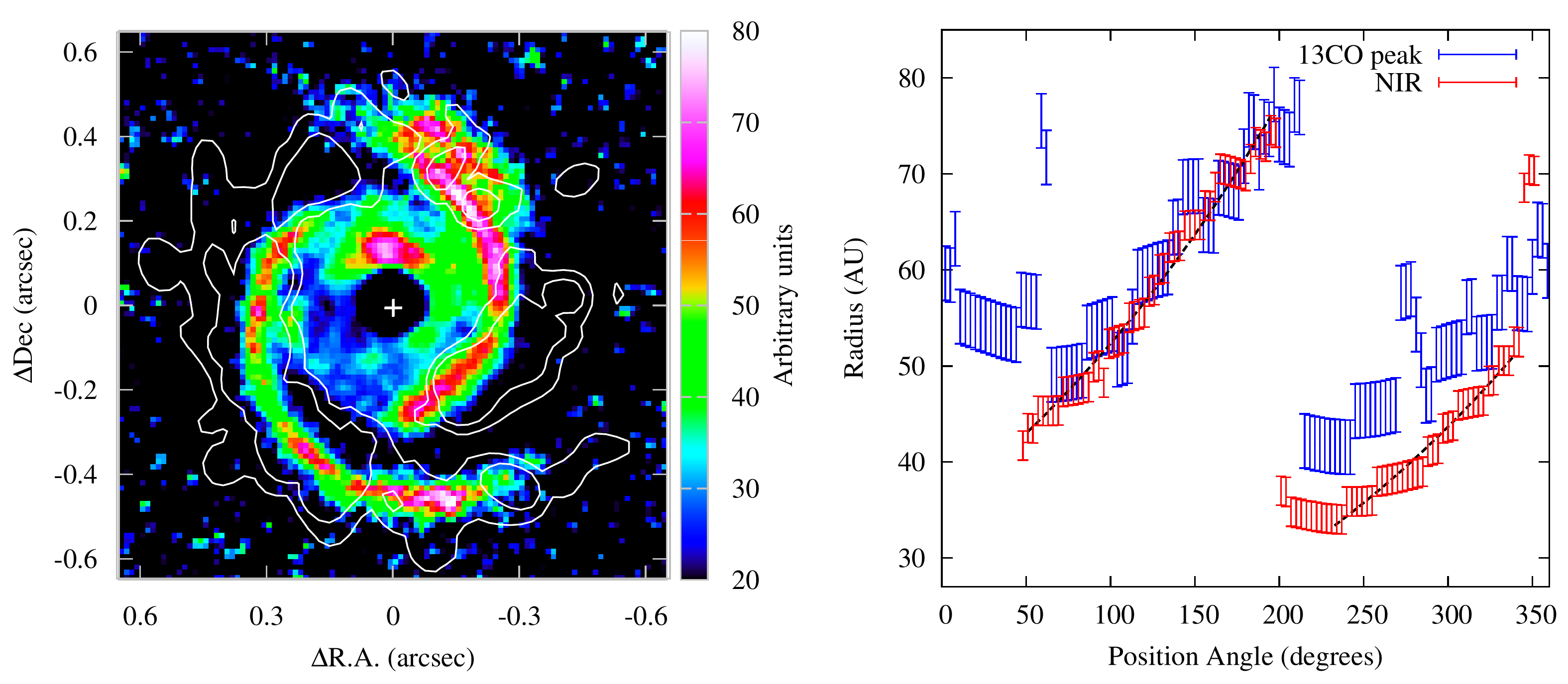}\\
  \caption{Left: map of the IR scattered light map, at an angular resolution of 0.027\arcsec\ , superimposed with white contours 
  at 45 and 50 K of the $^{13}$CO J = 3-2 peak emission, which is observed at an angular resolution is 0.23\arcsec\ $\times$ 
  0.18\arcsec. Right: crest of the $^{13}$CO emission in blue and of the near-IR scattered light emission in red. The fit of the 
  near-IR spirals is represented by black dashed lines.} 
  \label{Fig:overIRmom8}
\end{figure*}

The near-IR emission is expected to be optically thick and to be scattered efficiently by small dust grains of a few $\mu$m in 
protoplanetary disks. The polarized emission, therefore, traces fluctuations of the disk surface and has been used in 
observations to detect spirals or gap structures. Such observations of the disk around MWC~758 were made by 
\cite{Grad2013} using the Subaru telescope in near-IR via the $H$ Band (1.65 $\mu$m) in polarized emission and the 
$K$s Band (2.15 $\mu$m) in direct imaging. They detected 2 spirals, but with a larger contrast for the spiral starting on the east 
side of the star, and wrapping around the disk toward the south (SE spiral). \cite{Beni2015} redetected the spirals with 
VLT/SPHERE in scattered light in $Y$ Band  (1.04 $\mu$m) giving further insights on the disk surface. Here we propose 
to compare the observations of \cite{Beni2015} with our ALMA data to have a 3D view, of the surface and midplane, 
of the disk around MWC~758. 

In Figure~\ref{Fig:overIRmom8}, we compare the near-IR observations of \cite{Beni2015} with the $^{13}$CO peak emission. 
The $^{13}$CO peak emission is optically thick in the outer disk, with $\tau$ $\sim$ 3.7 and therefore is particularly 
sensitive to the disk temperature. We observe a clear correlation between the $^{13}$CO peak emission and the near-IR features, 
especially for the south-east spiral, suggesting that the spirals observed in near-IR are coincident with a local increase in 
temperature. The correlation with the north-west spiral is less evident. The spiral is mainly located at a smaller radius of 
40-50 au where the $^{13}$CO peak emission is more optically thin and depends also on the gas surface density, which increases up to 
60 au. Also, in the previous IR observations of \cite{Grad2013} and \cite{Beni2015}, the north-west spiral was less pronounced or 
azimuthally extended, suggesting a spiral less irradiated by the central star due to shadows cast from the inner regions 
of the disk or a lower enhancement in density and/or vertical elevation. Using a linear regression method, we fitted the position 
of the IR radial maxima by spirals, with a constant pitch angle, described by the following equation:
\begin{equation}
  R = R_0 ~ exp \left [tan(\alpha)(\theta-\theta_0) \right]
\end{equation}
with $\alpha$ the pitch angle, and $R_0$ the radial distance of the spiral at the azimuthal angle $\theta_0$. These fits are 
represented by dashed black lines in the right panel of Figure~\ref{Fig:overIRmom8}. The south-east 
spiral, fit at P.A. between 50$^\circ$ and 200$^\circ$, has a pitch angle of 12.7$^\circ$ and is located at 
$R_0$ = 43.0 au at P.A. = 50$^\circ$. The north-west spiral, fit between 230$^\circ$ and 340$^\circ$, features 
a similar pitch angle of 12.9$^\circ$, starting at the radial distance $R_0$ = 33.0 AU at P.A. = 230$^\circ$. The angular 
separation between the two spirals is of about 180$^\circ$, only varying slightly in the outer disk. There is no similar 
spiral-like components in the integrated emission of $^{13}$CO and C$^{18}$O, which are more optically thin and probe 
essentially the gas surface density.

\begin{figure*}
  \includegraphics[angle=-90,width=\textwidth]{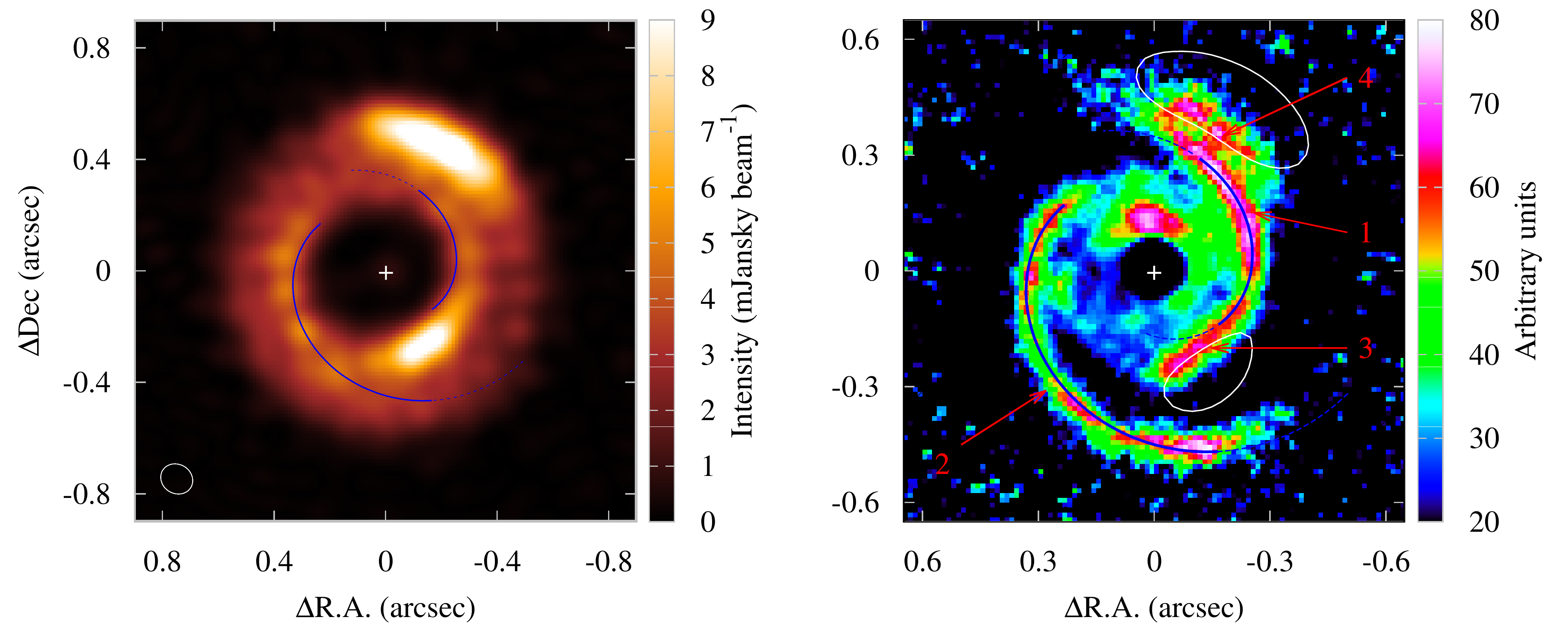}\\
  \caption{Left: Map of the dust emission with super-uniform weighting superimposed with the fit in blue solid lines of the 
  spirals observed in near-IR scattered light. The dotted blue lines show an extrapolation of the fit towards larger radius.  
  The angular resolution is 0.012\arcsec$\times$0.011\arcsec. Right: Map of the IR scattered light emission from \cite{Beni2015} 
  superimposed with white contours at 6.5 and 10 mJy beam$^{-1}$ of the continuum emission at 0.88 mm. The angular resolution 
  is 0.027\arcsec\ for the IR scattered light emission and 0.16\arcsec\ $\times$ 0.13\arcsec\ for the millimeter dust thermal 
  emission. The red numbers and arrows highlight the main features visible in IR in the outer ring.} 
  \label{Fig:su-IR}
\end{figure*}

We also compared the position of the infrared spirals with the dust emission obtained using super-uniform weighting in 
Figure~\ref{Fig:su-IR}. The fit of the spirals in near-IR scattered light is represented in blue solid lines and their 
extrapolation towards larger radii in blue dotted lines. In addition to the double-ring structure, the dust emission 
reveals two faint spirals located at slightly larger radius than the near-IR spirals (Shen, Tang et al., in prep.). Also, 
similarly to the near-IR observations, the pitch angle of the north-west spiral in the dust emission seems to increase at 
the proximity of the north dust clump. 

As already noticed by \cite{Beni2015}, and more recently by \cite{Regg2017} using the Keck telescope, the IR scattered light 
presents two other features in the outer ring, designated 3 and 4 in the right panel of Figure~\ref{Fig:su-IR}. They do not 
trace large spirals. The feature 3 has a negative pitch angle, and both have a very limited azimuthal extent. \cite{Regg2017} 
proposed that the north IR arc was possibly tracing the spiral 1, but at the bottom side of the disk. Our new observations show 
however that these two features are located at, and likely trace, the inner edge of the two dust clumps. Indeed, the local 
enhancements in dust surface density increase locally the absorption and scattering of the stellar irradiation at the inner 
edge of the dust clumps. This may then also rise up the temperature and the disk scale height, unveiling 
even more the two near-IR features. 

\vspace{1.5cm}

\section{Discussion}
\label{sec:discussion}

\subsection{Structure of the disk}
The disk around the Herbig star MWC 758 has a complex morphology which comprises a cavity, two rings, two large spirals and two 
dust clumps. According to our models, the disk, with a mass of about 1.3$\times$10$^{-3}$ M$_\odot$, is not massive enough to 
be gravitationally unstable \citep{Krat2016}. One explanation for such a complex structure is the presence of planet(s) which 
can interact with the protoplanetary disk and create features (annuli, spirals or vortices) which can be more easily detected 
than the planets themselves \citep{Kley2012, Baru2014}. 

We observe in the continuum emission two large dust clumps which have an elongated structure and an azimuthal/radial 
aspect ratio of approximatively 5 at our angular resolution. To our knowledge, MWC~758 is the only disk with V1247 Orionis, observed  
by \cite{Krau2017}, which features two dust clumps at two different radial distances. These structures are associated with azimuthal 
extensions dividing the disk in two rings, located at a distance from the star of about 57 and 81 au. By modeling the dust and 
gas emission, we conclude that the gas distribution is more diffused than the distribution of the large dust grains. Millimeter 
grains are gathered in compact regions with surface densities larger by a factor of 2.5 and 6.5 in the south and north dust 
clumps, respectively, compared to values encountered at other azimuthal angles. We do not detect increases of the 
gas surface density at the position of the dust clumps, what suggests a local augmentation of the dust-to-gas ratio, up to a 
value of $\sim$ 0.1.

The system MWC~758 is also famous for its two large and sharp spirals detected in near-IR scattered light by \cite{Grad2013}, 
\cite{Beni2015} and more recently by \cite{Regg2017}. These spirals feature a large pitch angle of $\sim$ 13$^\circ$ 
and are also observed with ALMA using the optically thick $^{13}$CO J = 3-2 peak emission and the optically thin continuum dust emission. 
This is the first time that spirals in protoplanetary disks are visible with these three tracers. The bright south-east spiral is 
observed at the same radius in IR and in $^{13}$CO. The IR scattered light traces the inner side of the spiral shock front at the 
disk surface, where the stellar emission is absorbed and locally heats the disk. In comparison, the spirals observed with the 
dust emission show fainter spiral structures and are located at a slightly larger radius, by a few au. The difference in the radial 
location might come from the width of the spirals, as we expect that the dust emission traces mainly the central position of the 
spirals. It can also be caused by the vertical propagation of the spirals which is expected to curl towards the central 
star at the disk surface instead of being perpendicular to the disk midplane \citep{Zhu2015}. 

The north-east spiral is less pronounced in the $^{13}$CO peak emission and less extended azimuhtally in IR scattered light 
\citep{Grad2013} than its south-west counterpart. It is possible that this spiral presents a lower scale height and/or density 
enhancement. We propose also that this might be due to a shadow cast by the inner regions of the disk. We detect a non-resolved 
source of continuum emission at the central star position, associated with a twist in the iso-velocity curves. Additionally, 
we detect an azimuthal minimum along the west direction in the $^{13}$CO and C$^{18}$O integrated emission and in the velocity 
dispersion, which supports the hypothesis of a mildly warped inner disk.

\subsection{Possible scenarios}

The origin of the spirals in MWC~758 has been under debate since their discovery by \cite{Grad2013}, and our ALMA 
observations also show the presence of two dust clumps and a dark ring, surrounding a large cavity. The south dust 
clump is located at the outer edge of the cavity, at 47 au, and the north dust clump, at the outer edge of the dark ring, at 
82 au. These regions present large gas surface density gradients where Rossby wave instabilities (RWI) are thought to develop 
and trigger vortices. The two dust clumps are also cospatial with the two azimuthal maxima detected in the continuum emission 
by \cite{Mari2015b} with the Very Large Array (VLA) at 33 GHz. This suggests that large dust grains are trapped into gas 
pressure maxima through aerodynamical drag, as already proposed by \cite{Mari2015b} for the north dust clump. Small local variations 
of 20-30 \% in the gas might not be visible in our observations but still be able to accrete efficiently dust particles 
\citep{Pini2012}. 

Many disks, like IRS 48 \citep{vdM2013}, AB Aur \citep{Tang2017}, SAO 206462 and SR 21 \citep{Pere2014}, among others, have shown 
large dust asymmetries that may also suit this scenario. As for MWC~758, most of these disks comprise a large cavity, which 
might be open by a stellar companion or planet(s). The probable detection by \cite{Regg2017} of a planet of 0.5-5 M$_{Jup}$ 
inside the cavity supports this idea. If the companion is massive enough, simulations of \cite{Ragu2017} have shown that it can also 
produce a large azimuthal variation of the gas surface density. This might be the case for the binary system HD~142527, which presents 
a stellar companion of about 0.25 M$_\odot$ and a large azimuthal ratio of $\sim$ 3.75 in the gas surface density \citep{Muto2015, Boeh2017}. 
This effect is however likely negligible in MWC~758 which has a low planet-to-star mass ratio, of about 0.001, and presents only 
moderate azimuthal variations of the gas surface density.

A scenario that might explain the complex structure of the outer regions of the MWC~758 system is the presence of a large 
planet at R $\geq$ 100 au. 3D hydrodynamical simulations of \cite{Zhu2015} and \cite{Dong2015} have shown that 
a massive planet of a few Jupiter mass in the outer regions would produce a pair of inner spiral arms, whose brightness and large 
pitch angle are compatible with the near-IR observations. Recent simulations of \cite{Bae2017} have also shown that, for a 
disk with low viscosity (i. e. $\alpha$ $\sim$ 10$^{-4}$), a massive planet can also produce a secondary gap in the inner regions 
of the planet orbit. This gap is linked to the secondary spiral and is located at approximatively 0.5-0.6 the orbital radius of the 
planet. This is consistent with the observation of the dark ring located between the two dust clumps. Such planet can also produce 
a tertiary spiral that might have been observed in the cavity of the disk by \cite{Regg2017}. However, this hypothetical planet has 
not been detected yet in the IR observations of \cite{Grad2013} and \cite{Regg2017}.

There are also processes that can explain part of the observations without the need of a planet in the outer regions, at R 
$\ge$ 100 au. A simple 
explanation for the dark ring is the presence of a supplementary planet of moderate mass inside the ring. It has also been proposed 
by \cite{Lobo2015} that a first vortex, formed at the edge of the central cavity, might accrete matter and cleans the orbit at a 
slightly larger radius. This process can then create a ring and be at the origin of a vortex of second generation. 
Nevertheless, these hypotheses alone do not explain the presence of the spirals. \cite{Mont2016} suggested that spirals might 
be induced dynamically as a consequence of a different irradiation from the star due to a warp in the inner regions of the disc. 
With the the possible presence of a warped inner disc in MWC~758 projecting shadows in the outer regions, this possibility has 
not to be discarded. Indeed, the two other disks which are known to feature an inner warped disk, HD~135344B and HD~142527, also 
have a large cavity and spirals \citep{Stok2016, Chri2014, Casa2015a}. It is also still unclear if a planet inside the cavity, with a 
non-coplanar or eccentric orbit, might produce such spirals. 

Finally, if massive enough, dust clumps might interact with, or 
even produce, spirals \citep{vdM2016b, Baru2016}. The spiral 
propagating outwards in the north direction crosses, and maybe starts from, the south dust clump. We also observe that the pitch 
angle of this spiral increases in the vicinity of the north dust clump, both in near-IR and in the dust emission. However, this 
scenario is not able to produce the south-east spiral which is not cospatial with any of the two dust clumps. It would 
also be surprising that the south clump is able to produce a large spiral, while the north dust clump, about 3 times more 
massive, is not.


\section{Conclusion}
\label{sec:conclu}

The transitional disk around the Herbig star MWC~758 was previously known to feature a large cavity empty of millimeter grains, 
two large spirals observed in IR polarized emission and at least one asymmetry in the dust distribution. We present new ALMA 
observations at an angular resolution of about 0.1-0.2\arcsec, which comprises dust emission at $\sim$ 0.88 millimeter, and 
the $^{13}$CO and C$^{18}$O J=3-2 emission lines. Our main findings about this system are:

\begin{enumerate}

  \item The outer disk features two large dust clumps, localized at 47 au and 82 au for the south and north clumps, 
  respectively. The local increase in the dust surface density is of $\sim$ 2.5 for the south clump and of $\sim$ 6.5 for 
  the north clump compared to values encountered at other azimuthal angles. The two dust concentrations are associated with 
  elongated structures which are tracing a double-ring structure. 
  
  \item The main hypothesis for such dust concentrations is the dust trapping scenario, where a local gas pressure/density 
  maximum aerodynamically accretes millimeter-size grains. The modeling of the gas and dust emission reveals smoother radial 
  variations in the gas densities than in the dust. We also did not find any specific gas concentration cospatial with the 
  dust clumps. This suggests lower local variations in the gas surface density, probably by no more than 20-30 \%. 
  Additionally, the dust clumps appear to be in eccentric positions compared to the underlying double ring 
  structure and the gas distribution, in regions with strong gradients in the surface density. 
  
  \item We observe the spirals, previously detected in near-IR scattered light, in the $^{13}$CO J = 3-2 peak emission lines 
  and in the continuum emission. The $^{13}$CO peak emission is optically thick and mainly traces variations of the disk 
  temperature while the dust emission is optically thin and probes the disk midplane. The spirals in IR and in the $^{13}$CO 
  J = 3-2 are located at the same radius while the spirals in continuum emission are located slightly at a larger radius. 
  This small offset in the radial location might be due to the spirals width and/or due to the vertical propagation of the 
  spirals, which might curl toward the central star at the disk surface.  
  
  \item Two arc-like features, designated 3 and 4 in Fig~\ref{Fig:su-IR}, were previously observed in near-infrared. Our 
  new ALMA observations show that these structures correspond to emission scattered at the inner edge of the two dust 
  clumps. 
  
  \item Dust emission at millimeter wavelengths is detected at the position of the central star and reveals the presence of 
  a small inner disk. The detection in the cavity of a twist in the velocity curves, as well as indications of a shadow 
  projected toward the west outer region, suggest that this disk is mildly warped.
  
  \item  Our observations are consistent with the existence of two massive planets. A planet in the inner region of the 
  disk which carves the cavity, and another planet in the outer regions to produce the spirals and the dark ring between the two 
  dust clumps. In this scenario, the dust vortices at the origin of the dust clumps are probably triggered by the Rossby wave 
  Instability due to the large radial gradients in the gas surface density. However, other possibilities exist and we discuss also 
  other processes that might be able to produce the features encountered in the disk.
  
\end{enumerate}

\begin{acknowledgements}
We acknowledge the anonymous referee for constructive comments and useful suggestions that improved this paper.
Y.B. and A.I. acknowledge support from the NASA Origins of Solar Systems program through award NNX15AB06G. A.I. acknowledges support 
from the NSF Grant No. AST-1535809. E.W. acknowledges support from the NRAO Student Observing Support Grant No. AST-0836064. M.B. 
acknowledges funding from ANR of France under contract number ANR-16-CE31-0013 (Planet Forming Disks). J.M.C. 
acknowledges support from the National Aeronautics and Space Administration under Grant No. 15XRP15 20140 issued through 
the Exoplanets Research Program. Y-W. T. is supported by the Ministry of Science and Technology (MoST) in Taiwan through grant 
MoST 103-2119-M-001-010-MY2. This paper makes use of the following ALMA data 
\dataset[ADS/JAO.ALMA2012.1.00725.S]{https://almascience.nrao.edu/aq/?project_code=2012.1.00725.S}. ALMA is a partnership of ESO 
(representing its member states), NSF 
(USA) and NINS (Japan), together with NRC (Canada) and NSC and ASIAA (Taiwan) and KASI (Republic of Korea), in cooperation with the 
Republic of Chile. The Joint ALMA Observatory is operated by ESO, AUI/NRAO and NAOJ. The National Radio Astronomy  Observatory is a 
facility of the National Science Foundation operated under cooperative agreement by Associated Universities, Inc. This work has also 
made use of data from the European Space Agency (ESA) mission {\it Gaia}(\url{http://www.cosmos.esa.int/gaia}), 
processed by the {\it Gaia} Data Processing and Analysis Consortium (DPAC, \url{http://www.cosmos.esa.int/web/gaia/dpac/consortium}). 
Funding for the DPAC has been provided by national institutions, in particular the institutions participating in the {\it Gaia} 
Multilateral Agreement. The description of the Gaia mission is described in \cite{Gaiab2016} and the data release 1 in \cite{Gaiaa2016}.
\end{acknowledgements}





\begin{thebibliography}{99}     
   
\bibitem[ALMA Partnership et al.(2015)]{ALMA2015} ALMA Partnership, Brogan, C.~L., P{\'e}rez, L.~M., et al.\ 2015, \apjl, 808, L3    
\bibitem[Anders \& Grevesse(1989)]{Ande1989} Anders, E., \& Grevesse, N.\ 1989, \gca, 53, 197    
\bibitem[Andrews et al.(2011)]{Andr2011} Andrews, S.~M., Wilner, D.~J., Espaillat, C., et al.\ 2011, \apj, 732, 42   
\bibitem[Bae et al.(2017)]{Bae2017} Bae, J., Zhu, Z., \& Hartmann, L.\ 2017, \apj, 850, 201 
\bibitem[Baruteau et al.(2014)]{Baru2014} Baruteau, C., Crida, A., Paardekooper, S.-J., et al.\ 2014, Protostars and Planets VI, 667 
\bibitem[Baruteau \& Zhu(2016)]{Baru2016} Baruteau, C., \& Zhu, Z.\ 2016, \mnras, 458, 3927 
\bibitem[Benisty et al.(2015)]{Beni2015} Benisty, M., Juhasz, A., Boccaletti, A., et al.\ 2015, \aap, 578, L6 
\bibitem[Boehler et al.(2017)]{Boeh2017} Boehler, Y., Weaver, E., Isella, A., et al.\ 2017, \apj, 840, 60 
\bibitem[Casassus et al.(2015a)]{Casa2015a} Casassus, S., Marino, S., P{\'e}rez, S., et al.\ 2015a, \apj, 811, 92 
\bibitem[Casassus et al.(2015b)]{Casa2015b} Casassus, S., Wright, C.~M., Marino, S., et al.\ 2015b, \apj, 812, 126 
\bibitem[Christiaens et al.(2014)]{Chri2014} Christiaens, V., Casassus, S., Perez, S., van der Plas, G., \& M{\'e}nard, 
F.\ 2014, \apjl, 785, L12 
\bibitem[Dipierro et al.(2016)]{DiPi2016} Dipierro, G., Laibe, G., Price, D.~J., \& Lodato, G.\ 2016, \mnras, 459, L1 
\bibitem[Dong et al.(2015)]{Dong2015} Dong, R., Zhu, Z., Rafikov, R.~R., \& Stone, J.~M.\ 2015, \apjl, 809, L5 
\bibitem[Dong et al.(2017)]{Dong2017} Dong, R., van der Marel, N., Hashimoto, J., et al.\ 2017, \apj, 836, 201 
\bibitem[Dullemond et al.(2012)]{Dull2012} Dullemond, C.~P., Juhasz, A., Pohl, A., et al.\ 2012, Astrophysics Source Code Library, ascl:1202.015
\bibitem[Eisner et al.(2004)]{Eisn2004} Eisner, J.~A., Lane, B.~F., Hillenbrand, L.~A., Akeson, R.~L., \& Sargent, A.~I.\ 2004, \apj, 613, 1049 
\bibitem[Facchini et al.(2018)]{Facc2018} Facchini, S., Juh{\'a}sz, A., \& Lodato, G.\ 2018, \mnras, 473, 4459
\bibitem[Fedele et al.(2017)]{Fede2017} Fedele, D., Carney, M., Hogerheijde, M.~R., et al.\ 2017, \aap, 600, A72 
\bibitem[Gaia Collaboration et al.(2016a)]{Gaiaa2016} Gaia Collaboration, Brown, A.~G.~A., Vallenari, A., et al.\ 2016, \aap, 595, A2 
\bibitem[Gaia Collaboration et al.(2016b)]{Gaiab2016} Gaia Collaboration, Prusti, T., de Bruijne, J.~H.~J., et al.\ 2016, \aap, 595, A1 
\bibitem[Gonzalez et al.(2015)]{Gonz2015} Gonzalez, J.-F., Laibe, G., Maddison, S.~T., Pinte, C., \& M{\'e}nard, F.\ 2015, \mnras, 454, L36 
\bibitem[Grady et al.(2013)]{Grad2013} Grady, C.~A., Muto, T., Hashimoto, J., et al.\ 2013, \apj, 762, 48 
\bibitem[Guilera \& S{\'a}ndor(2017)]{Guil2017} Guilera, O.~M., \& S{\'a}ndor, Z.\ 2017, \aap, 604, A10 
\bibitem[Isella et al.(2008)]{Isel2008} Isella, A., Tatulli, E., Natta, A., \& Testi, L.\ 2008, \aap, 483, L13 
\bibitem[Isella et al.(2010)]{Isel2010} Isella, A., Natta, A., Wilner, D., Carpenter, J.~M., \& Testi, L.\ 2010, \apj, 725, 1735 
\bibitem[Isella et al.(2016)]{Isel2016} Isella, A.,Guidi, G., Testi, L., et al.\ 2016, Phys. Rev. Lett., 117, 25
\bibitem[Johansen et al.(2009)]{Joha2009} Johansen, A., Youdin, A., \& Klahr, H.\ 2009, \apj, 697, 1269 
\bibitem[Kley \& Nelson(2012)]{Kley2012} Kley, W., \& Nelson, R.~P.\ 2012, \araa, 50, 211 
\bibitem[Kratter \& Lodato(2016)]{Krat2016} Kratter, K., \& Lodato, G.\ 2016, \araa, 54, 271 
\bibitem[Kraus et al.(2017)]{Krau2017} Kraus, S., Kreplin, A., Fukugawa, M., et al.\ 2017, \apjl, 848, L11 
\bibitem[Lesur \& Papaloizou(2010)]{Lesu2010} Lesur, G., \& Papaloizou, J.~C.~B.\ 2010, \aap, 513, A60 
\bibitem[Lobo Gomes et al.(2015)]{Lobo2015} Lobo Gomes, A., Klahr, H., Uribe, A.~L., Pinilla, P., \& Surville, C.\ 2015, \apj, 810, 94 
\bibitem[Marino et al.(2015a)]{Mari2015a} Marino, S., Perez, S., \& Casassus, S.\ 2015a, \apjl, 798, L44 
\bibitem[Marino et al.(2015b)]{Mari2015b} Marino, S., Casassus, S., Perez, S., et al.\ 2015b, \apj, 813, 76 
\bibitem[McMullin et al.(2007)]{Mcmu2007} McMullin, J.~P., Waters, B., Schiebel, D., Young, W., \& Golap, K.\ 2007, Astronomical Data 
Analysis Software and Systems XVI, 376, 127 
\bibitem[Montesinos et al.(2016)]{Mont2016} Montesinos, M., Perez, S., Casassus, S., et al.\ 2016, \apjl, 823, L8 
\bibitem[Muto et al.(2015)]{Muto2015} Muto, T., Tsukagoshi, T., Momose, M., et al.\ 2015, \pasj, 67, 122 
\bibitem[Nelson et al.(2013)]{Nels2013} Nelson, R.~P., Gressel, O., \& Umurhan, O.~M.\ 2013, \mnras, 435, 2610 
\bibitem[Owen et al.(2011)]{Owen2011} Owen, J.~E., Ercolano, B., \& Clarke, C.~J.\ 2011, \mnras, 412, 13 
\bibitem[P{\'e}rez et al.(2014)]{Pere2014} P{\'e}rez, L.~M., Isella, A., Carpenter, J.~M., \& Chandler, C.~J.\ 2014, \apjl, 783, L13 
\bibitem[P{\'e}rez et al.(2016)]{Pere2016} P{\'e}rez, L.~M., Carpenter, J.~M., Andrews, S.~M., et al.\ 2016, Science, 353, 1519 
\bibitem[Pinilla et al.(2012)]{Pini2012} Pinilla, P., Birnstiel, T., Ricci, L., et al.\ 2012, \aap, 538, A114 
\bibitem[Pinilla et al.(2015)]{Pini2015} Pinilla, P., de Juan Ovelar, M., Ataiee, S., et al.\ 2015, \aap, 573, A9
\bibitem[Qi et al.(2011)]{Qi2011} Qi, C., D'Alessio, P., {\"O}berg, K.~I., et al.\ 2011, \apj, 740, 84 
\bibitem[Quillen et al.(2005)]{Quil2005} Quillen, A.~C., Varni{\`e}re, P., Minchev, I., \& Frank, A.\ 2005, \aj, 129, 2481 
\bibitem[Ragusa et al.(2017)]{Ragu2017} Ragusa, E., Dipierro, G., Lodato, G., Laibe, G., \& Price, D.~J.\ 2017, \mnras, 464, 1449 
\bibitem[Reg{\'a}ly et al.(2012)]{Rega2012} Reg{\'a}ly, Z., Juh{\'a}sz, A., S{\'a}ndor, Z., \& Dullemond, C.~P.\ 2012, \mnras, 419, 1701 
\bibitem[Reggiani et al.(2017)]{Regg2017} Reggiani, M., Christiaens, V., Absil, O., et al.\ 2017, arXiv:1710.11393 
\bibitem[Ruge et al.(2016)]{Ruge2016} Ruge, J.~P., Flock, M., Wolf, S., et al.\ 2016, \aap, 590, A17 
\bibitem[Stolker et al.(2016)]{Stok2016} Stolker, T., Dominik, C., Avenhaus, H., et al.\ 2016, \aap, 595, A113 
\bibitem[Tang et al.(2017)]{Tang2017} Tang, Y.-W., Guilloteau, S., Dutrey, A., et al.\ 2017, \apj, 840, 32 
\bibitem[van den Ancker et al.(1998)]{vdA1998} van den Ancker, M.~E., de Winter, D., \& Tjin A Djie, H.~R.~E.\ 1998, \aap, 330, 145 
\bibitem[van der Marel et al.(2013)]{vdM2013} van der Marel, N., van Dishoeck, E.~F., Bruderer, S., et al.\ 2013, Science, 340, 1199 
\bibitem[van der Marel et al.(2015)]{vdM2015} van der Marel, N., van Dishoeck, E.~F., Bruderer, S., P{\'e}rez, L., \& Isella, A.\ 2015, \aap, 579, A106 
\bibitem[van der Marel et al.(2016a)]{vdM2016a} van der Marel, N., van Dishoeck, E.~F., Bruderer, S., et al.\ 2016a, \aap, 585, A58 
\bibitem[van der Marel et al.(2016b)]{vdM2016b} van der Marel, N., Cazzoletti, P., Pinilla, P., \& Garufi, A.\ 2016b, \apj, 832, 178 
\bibitem[van Leeuwen(2007)]{vanL2007} van Leeuwen, F.\ 2007, \aap, 474, 653 
\bibitem[Zhang et al.(2015)]{Zhan2015} Zhang, K., Blake, G.~A., \& Bergin, E.~A.\ 2015, \apjl, 806, L7 
\bibitem[Zhu et al.(2015)]{Zhu2015} Zhu, Z., Dong, R., Stone, J.~M., \& Rafikov, R.~R.\ 2015, \apj, 813, 88 

\end{thebibliography}
\end{document}